\documentclass[aps,prb,twocolumn,groupedaddress,floatfix]{revtex4-2}

\usepackage{graphicx}
\usepackage{dcolumn}
\usepackage{bm}
\usepackage{mathtools}

\usepackage{physics}

\newlength\mylen  \setcitestyle{citesep={,\kern-\mylen}}

\usepackage{xcolor}

\begin{document}

\title{Heavy-Hole Spin Relaxation in Quantum Dots: Isotropic versus Anisotropic Effects }
\author{
Dalton Forbes$^1$, Sanjay Prabhakar$^{1a)}$, Ruma De$^1$, Himadri S. Chakraborty$^1$,  Roderick Melnik$^2$
}
\affiliation{
Department of Natural Sciences, Dean L. Hubbard Center for Innovation, Northwest Missouri State University, 800 University Drive, Maryville, MO 64468\\
$^2$MS2Discovery Interdisciplinary Research Institute, Wilfrid Laurier University, 75 University Ave W, Waterloo, Ontario, Canada N2L 3C5
}

\affiliation{
$^{a)}$Author to whom correspondence should be addressed: sanjay@nwmissouri.edu}

\date{June 21, 2023}

\begin{abstract}
Non-charge based logic in single-hole spin of semiconductor quantum dots (QDs) can be controlled by anisotropic gate potentials providing a notion for making next generation solid-state quantum devices. In this study, we investigate the isotropic and anisotropic behavior of phonon mediated spin relaxation of heavy-hole spin hot spots in QDs. For the electron spin in isotropic QDs, hot spots are known to be always present due to the Rashba spin-orbit coupling. But for heavy holes in isotropic dots, we show that the occurrences of spin hot spots are sensitive to the bulk g-factor: the hot spot for Rashba coupling in InAs and GaSb dots arises because these materials possess negative bulk g-factor, while that for the Dresselhaus coupling in GaAs and InSb dots is found due to their positive bulk g-factor. For anisotropic QDs, on the other hand, the spin hot spot is universally present due to their broken in-plane rotational symmetry. Further, the increasing electric field, that strengthens the Rashba coupling, is shown to cover a wide range of magnetic field by the hot spots. Results demonstrate that the magnetic field, choice of dot materials and size anisotropy can act as effective control parameters which can be experimentally used to design the device for detecting the phonon mediated heavy-hole spin-relaxation behavior of III-V semiconductor QDs.
\end{abstract}

\maketitle

\section{Introduction}

Manipulation of a single-hole spin with gate-controlled electric fields, magnetic fields, as well as optical pumping methods in confined nanostructure-based complementary metal-oxide semiconductor (CMOS) devices is an ongoing proposal for the solid-state realization of quantum computing and quantum information processing applications~\cite{gerardot08,pribiag13,greilich11,trif09,fischer08,hendrickx20,pericles20,vecchio23}. In these metal-oxide field effect transistors (MOSFET), it is possible that heavy-hole spins can be initialized from one of the source spin currents and can be detected from drain spin currents with the application of gate controlled electric field through the gate oxide layer~\cite{prabhakar09,alhasani22,li23,fernandez22,li23a,shalak23,kepa23,lei23}. When a spin current is chosen as a qubit, its decay time is given by a spin-relaxation time or decoherence time $T_2\approx 2T_1$, where $T_1$ is the hole relaxation time~\cite{bulaev05}. To preserve the quantum behavior of a qubit, the relaxation time should be several orders of magnitude longer than the minimum time required to initialize and read out the spin currents~\cite{zhang18,rotta17,bao22,delaney22}. Large decoherence times of nanostructure qubits have been reported experimentally and theoretically~\cite{hendrickx20,burkard23,camenzind22,adelsberger22,lawrie20,bulaev05}. At low temperature measurements due to the interaction of spin-orbit coupling with the phonon, the spin hot spot, i.e. the strong admixture of spin states to other states, is observed in the decoherence of electron and hole spins in quantum dots (QDs)~\cite{bulaev05a,prabhakar13}. In III-V semiconductor, the  spin-orbit coupling is mainly dominated by the Rashba coupling and linear Dresselhaus coupling~\cite{bulaev05a,prabhakar13}. The Rashba coupling arises due to the structural inversion asymmetry along the growth direction while the Dresselhaus coupling is due to the bulk inversion asymmetry of the crystal lattice.

In semiconductor nanostructures including QDs, gate control of single-electron and single-hole spin degree of freedom is an important recipe for building spin qubits~\cite{burkard23,bosco22,abadillo23,spethmann22,dykman23,prabhakar13,prabhakar14a}. Both isotropic and anisotropic QDs are potential candidates for the realization of spin qubits or controlling their spin behavior for applications in solid state realization of quantum computing. The spin hot spot that can be observed in these devices significantly reduces the decoherence time due to the very strong mixing of spin states with the other available states. This may be considered a drawback but may also be useful for quantum entanglement~\cite{collard19,brooks23,niegemann22}. Hence, finding an ideal location of the spin hot spot in both isotropic and anisotropic dots is an important ingredient to initialize and to read out spin qubits. In isotropic QDs, it is well known that the electron spin hot spot is observed in the phonon mediated spin relaxation for the case of pure Rashba coupling, whereas it is entirely absent for the Dresselhaus case~\cite{prabhakar13,bulaev05a}. The electron spin hot spot is always present in anisotropic QDs due to the broken in-plane rotational symmetry~\cite{prabhakar13}. However, behaviors of hot spot for the heavy-hole spin in both isotropic and anisotropic QDs have not yet been explored in details.

In this paper, we show that the heavy-hole spin hot spot in phonon mediated spin relaxation is very sensitive to the bulk g-factor of the heavy hole in QDs. More precisely, in GaAs and InSb isotropic dots, the spin hot spot for a heavy hole can be observed for the pure Dresselhaus spin-orbit coupling due to the presence of a positive bulk g-factor of the hole. In contra-distinction, for each of GaSb and InAs isotropic dots, a spin hot spot is predicted for the pure Rashba coupling because these materials possess a negative bulk g-factor. However, the anisotropy in the dot shape breaks the in-plane rotational symmetry. As a result, the spin hot spot in the phonon mediated heavy-hole spin relaxation can appear in both pure Rashba and Dresselhaus spin-orbit couplings for anisotropic dots, akin to the electron spin case. The rotational symmetry of heavy-hole QDs can be broken by inducing the anisotropy through the design of external gates. Furthermore, at low magnetic fields and for small lateral sizes of the dots the pure Dresselhaus case suggests an asymmetric behavior of heavy-hole spin relaxation: the relaxation rate increases until it reaches a maximum, then decreases and again increases until the level-crossing of the states occurs. As the strength of the Rashba spin-orbit coupling increases with the rise of electric field, the spin hot spot is shown to cover a wide range of magnetic fields.

The paper is organized as follows. In section~\ref{theoretical-model}, we develop a theoretical model for isotropic and anisotropic heavy-hole spin relaxation mediated by phonon that will allow us to investigate the interplay between the Rashba and the linear Dresselhaus spin-orbit couplings. In section~\ref{computational-method}, we briefly describe the computational diagonalization technique of finite element method simulations to find the energy spectrum and the matrix elements of the phonon mediated spin relaxation rate in QDs. In section~\ref{results-and-discussions}, we present and discuss the results of isotropic and anisotropic heavy-hole spin-relaxation rates versus the magnetic field and the QD radius for the pure Dresselhaus and the mixed Rashba-Dresselhaus (RD) coupling cases in III-V semiconductor materials of zinc blend GaAs, GaSb, InAs and InSb QDs. Finally, section~\ref{conclusion} summarizes our results.

\section{Theoretical Model}\label{theoretical-model}

For low temperature measurements, it is possible to decouple heavy-hole states from light-hole and spin split-off states in the Luttinger-Kohn Hamiltonian~\cite{bulaev05,meier12,luttinger55,prabhakar13coupled}
\begin{equation}
H = H_{LK} -\frac{\gamma}{\eta} \mathbf{J} \mathbf{\cdot} \mathbf{\Omega}, \label{LK}
\end{equation}
where $H_{LK}$ is the Luttinger-Kohn Hamiltonian. The contribution of $\gamma$ comes from the bulk inversion symmetry, $\eta$ is the spin split-off energy, $\mathbf{J}=\left(J_x, J_y, J_z\right)$ are $4\times 4$ matrices corresponding to spin 3/2, and $\mathbf{\Omega}=\left(\Omega_x, \Omega_y, \Omega_z\right)$ are momentum operators. Here, $\Omega_x = P_x\left(P_y^2-P_z^2\right)$, $\Omega_y = P_y\left(P_z^2-P_x^2\right)$, $\Omega_z = P_z\left(P_x^2-P_y^2\right)$. We assume that the heavy-hole and light-hole splittings and spin split-off are large in such a way that at low temperature measurements we can use a two-band Kane model for the heavy hole in the presence of a magnetic field in z-direction as~\cite{bulaev05,meier12,luttinger55,prabhakar13coupled}
\begin{eqnarray}
  H &=& H_0 + H_R + H_D, \label{H} \\
  H_0 &=& \frac{1}{2m}\left(P_x^2+P_y^2\right) + \frac{1}{2}m \omega_0^2 \left(a x^2 + b y^2\right) + \frac{\Delta}{2}  \sigma_z, ~~~ \label{H0} \\
 H_R &=& i\alpha_R\left(\sigma_+P_-^3-\sigma_-P_+^3\right),\label{HR}\\
  H_D &=& -\alpha_D\left(\sigma_+P_-P_+P_-+\sigma_-P_+P_-P_+\right).\label{HD}
\end{eqnarray}
Here, $\mathbf{P}=p + e/\mathbf{A}$ with $\mathbf{A}=B \left(-y \sqrt b ,x\sqrt a,0\right)/\left(\sqrt a+\sqrt b\right)$ is the vector potential, $\mathbf{p}$ is the momentum operator,  and the 2nd term in Eq.\,(\ref{H0}) can be used to control the shape and size of the QD [e.g., for an isotropic QD ($a=b$) and for an anisotropic QD (a$\neq$b)].  The 3rd term in Eq.\,(\ref{H0}) is the Zeeman spin-splitting energy due to the applied magnetic field $B$ in z-direction, where $\Delta=g_{hh}\mu_B B$. Here, $m$ is the effective mass of the heavy hole in QDs, $\ell_0 = \sqrt{\hbar/m\omega_0}$ is the radius of the lateral size of the dots and $g_{hh}$ is the bulk g-factor of the heavy hole. The spin-orbit coupling, $H_{so} = H_R+H_D$, where $H_R$ is the Rashba and $H_D$ is the Dresselhaus spin-orbit coupling.
Here $\alpha_R=3\gamma_0\gamma_R E/2m_0\Delta$ and $\alpha_D=3\gamma_0\gamma_D\hbar^2\kappa^2/2m_0\eta\Delta$ are the coefficients of Rashba and Dresselhaus couplings, $\kappa=\left(2meE/\hbar^2\right)^{1/3}$ is the thickness of a two-dimensional hole gas, $E$ is the applied electric field along z-direction, $\sigma_{\pm}=\sigma_x\pm i \sigma_y$, where $\sigma_x$, $\sigma_y$ are the Pauli matrices, and $P_{\pm}=P_x\pm i P_y$. The energy bands for unperturbed Hamiltonian, $H_0$, can be written as
\begin{equation}
\varepsilon^0_{n_+,n_-,\pm}=\left(n_++n_-+1\right)\hbar\omega_++\left(n_+-n_-\right)\hbar\omega_-\pm \frac{\Delta}{2}, \label{epsilon-a}\\
\end{equation}
where $\omega_{\pm}=\frac{1}{2}\left[\omega_c^2+\omega_0^2\left(\sqrt a\pm\sqrt b\right)^2\right]^{1/2}$, and $n_{\pm}$ are the eigenvalues of the Fock-Darwin number operators $a_{\pm}^\dagger a_{\pm}$. Here, $a_{\pm}$ and $a_{\pm}^\dagger$ are usual annihilation and creation operators. Also, we label the Fock-Darwin states as $|n_+, n_-,\pm\rangle$ with $\pm$ being the eigenvalues of the Pauli spin matrix along z-direction~\cite{prabhakar13}. In terms of raising and lowering operators ($a_{\pm}$ and $a^\dagger_{\pm}$) for symmetric QDs (a=b=1), we can write the spin-orbit coupling terms as,
\begin{widetext}
\begin{equation}
H_R = -\alpha \sigma_+
\begin{bmatrix}
\left\{ \left(\frac{\hbar}{\ell}\right)^3 + \frac{3}{2} \frac{eB\hbar^2}{\ell}  + \frac{3}{4}\left(eB\right)^2\hbar\ell + \left(\frac{eB\ell}{2}\right)^3 \right\} a_+^3 +
\left\{ -3\left(\frac{\hbar}{\ell}\right)^3 - \frac{3}{2} \frac{eB\hbar^2}{\ell} + \frac{3}{4}\left(eB\right)^2 \hbar\ell+ 3 \left(\frac{eB\ell}{2}\right)^3 \right\} a_+^2 a_-^\dagger + \\
\left\{ 3\left(\frac{\hbar}{\ell}\right)^3 - \frac{3}{2} \frac{eB\hbar^2}{\ell} - \frac{3}{4}\left(eB\right)^2 \hbar\ell+ 3 \left(\frac{eB\ell}{2}\right)^3 \right\} a_+ {a_+^\dagger}^2
+ \left\{ -\left(\frac{\hbar}{\ell}\right)^3 + \frac{3}{2} \frac{eB\hbar^2}{\ell} - \frac{3}{4}\left(eB\right)^2 \hbar\ell+  \left(\frac{eB\ell}{2}\right)^3 \right\} {a_-^\dagger}^3
\end{bmatrix}
+H.c., \label{HR-1}
\end{equation}
\end{widetext}
\begin{widetext}
\begin{equation}
H_D = -i\beta \sigma_+
\begin{bmatrix}
\left(\frac{\hbar}{\ell}\right)^3\left\{  a_+^2a_- -2a_+ -n_+a_+ -2 a_+n_- + a_-^\dagger + 2 n_+ a_-^\dagger + n_-a_-^\dagger - a_+^\dagger {a_-^\dagger}^2\right\} +\\
\frac{1}{2}\left(\frac{eB\hbar^2}{\ell}\right)^3\left\{  a_+^2a_- -2a_+ -3n_+a_+ +2 a_+n_- + a_-^\dagger + 2 n_+ a_-^\dagger -3 n_-a_-^\dagger + a_+^\dagger {a_-^\dagger}^2 \right\}  + \\
\frac{1}{4}\left(eB\right)^2 \hbar \ell \left\{  a_+^2a_- +2a_+ +3n_+a_+ -2 a_+n_- + a_-^\dagger + 2 n_+ a_-^\dagger - 3 n_-a_-^\dagger - a_+^\dagger {a_-^\dagger}^2 \right\}  +\\
\left(\frac{eB\ell}{2}\right)^3   \left\{ a_+^2a_- +2a_+ +n_+a_+ +2 a_+n_- + a_-^\dagger + 2 n_+ a_-^\dagger + n_-a_-^\dagger + a_+^\dagger {a_-^\dagger}^2 \right\}
\end{bmatrix}
+H.c., \label{HD-1}
\end{equation}
\end{widetext}
where H.c.\ represents the Hermitian conjugate, $\ell=\sqrt{\hbar/m\Omega}$ is the hybrid orbital length and $\Omega=\sqrt{\omega_0^2+\omega_c^2/4}$ for isotropic QDs. Operators algebra for anisotropic QDs is extremely lengthy and we only perform computational calculations of full Hamiltonian of~(\ref{H}).

We now turn to the calculation of the phonon induced spin-relaxation rate at low temperatures in QDs. Since we deal with small energy transfers, we consider the interaction of the piezophonon with holes to write as~\cite{gantmakher-book,prabhakar13,khaetskii00,khaetskii01},
\begin{equation}
u^{\mathbf{q}\alpha}_{ph}\left(\mathbf{r},t\right)=\sqrt{\frac{\hbar}{2\rho V \omega_{\mathbf{q}\alpha}}} e^{i\left(\mathbf{q\cdot r} -\omega_{q\alpha} t\right)} e A_{\mathbf{q}\alpha}b^{\dag}_{\mathbf{q}\alpha} + H.c.,
\label{u}
\end{equation}
where $\rho$ is the crystal mass density, $V$ is the volume of the QD.  Also, $b^{\dag}_{\mathbf{q}\alpha}$ creates an acoustic phonon with wave vector $\mathbf{q}$ and polarization $\hat{e}_\alpha$, where $\alpha=l,t_1,t_2$ are chosen as one longitudinal and two transverse modes of the induced phonon in the dots. In Eq.\,(\ref{u}), $A_{\mathbf{q}\alpha}=\hat{q}_i\hat{q}_k e\beta_{ijk} e^j_{\mathbf{q}\alpha}$ is the amplitude of the electric field created by the phonon strain, where $\hat{\mathbf{q}}=\mathbf{q}/q$ and $e\beta_{ijk}=eh_{14}$ for $i\neq k, i\neq j, j\neq k$. The polarization directions of the induced phonon are $\hat{e}_l=\left(\sin\theta \cos\phi, \sin\theta \sin\phi, \cos\theta \right)$, $\hat{e}_{t_1}=\left(\cos\theta \cos\phi, \cos\theta \sin\phi, -\sin\theta \right)$ and $\hat{e}_{t_2}=\left(-\sin\phi, \cos\phi, 0 \right)$. Note that the Gaussian cutoff at $\mathbf{q}\sim  1/\ell,$ with $\ell$ being the hybrid orbital length of the dots, restricts the coupling of long wavelength phonon to the dots so the frequency of phonon can be
replaced by its value at the center of the Brillouin zone (dispersonless approximation)~\cite{gantmakher-book,khaetskii00,khaetskii01}.  Based on the Fermi Golden Rule, the phonon induced spin-relaxation rate in the QDs is given by~\cite{prabhakar13,khaetskii01}
\begin{equation}
w_0=\frac{2\pi}{\hbar}\int \frac{d^3\mathbf{q}}{\left(2\pi\right)^3}\sum_{\alpha=l,t}\arrowvert M\left(\mathbf{q}\alpha\right)\arrowvert^2\delta\left(\hbar s_\alpha \mathbf{q}-\varepsilon_{f}+\varepsilon_{i}\right),
\label{1-T1}
\end{equation}
where  $s_l$, $s_t$ are the longitudinal and transverse acoustic phonon velocities in QDs.  The matrix element $M\left(\mathbf{q}\alpha\right)=\langle \psi_i|u^{\mathbf{q}\alpha}_{ph}\left(\mathbf{r},t\right)|\psi_f\rangle$ with the emission of one phonon $\mathbf{q}\alpha$ has been calculated perturbatively and numerically~\cite{khaetskii01,prabhakar13,comsol}. Here $|\psi_i\rangle$ and $|\psi_f\rangle$ correspond to the initial and final states of the Hamiltonian $H$. More precisely, we write the dipole matrix element as,
\begin{eqnarray}
M\left(\mathbf{q}\alpha\right)=\sqrt{\frac{\hbar}{2\rho \omega_{q\alpha}}}
\langle \psi_i
\begin{vmatrix}
  1-i\mathbf{q\cdot r}-\frac{1}{2}\left(\mathbf{q\cdot r}\right)^2 \\
  +\frac{i}{6}\left(\mathbf{q\cdot r}\right)^3+\cdots
\end{vmatrix}|
\psi_f\rangle.~~~
\label{Mq}
\end{eqnarray}
As can be seen from Eq.\,(\ref{HR-1}), the state $|0,0\rangle$ interacts with $|0,3\rangle$ and $|3,0\rangle$ due to the Rashba spin-orbit coupling. To capture such an interaction, we expand the hole-phonon coupling beyond the dipole approximation so that we can calculate the influence of the Rashba coupling in the phonon mediated spin relaxation of heavy hole. Similarly, from Eq.\,(\ref{HD-1}), the state $|0,0\rangle$ interacts with $|1,0\rangle$, $|0,1\rangle$, $|2,1\rangle$ and $|1,2\rangle$ due to the Dresselhaus spin-orbit coupling. Hence, within the dipole approximation, we find the spin relaxation due to the interaction of the states $|0,0\rangle$ with $|1,0\rangle$ and $|0,1\rangle$ for the pure Dresselhaus case. Also, going beyond the dipole approximation allows us to find the spin relaxation due to the interaction of the state $|0,0\rangle$ with $|2,1\rangle$ and $|1,2\rangle$ for the Dresselhaus case. In other words, in Eq.~(\ref{Mq}),  $\langle \psi_i|\mathbf{q\cdot r}|\psi_f\rangle$ and $\langle \psi_i|\left(\mathbf{q\cdot r}\right)^3|\psi_f\rangle$ induce the non-vanishing matrix elements. Within the dipole approximation, the spin-relaxation rate due to the interaction of states $|0,0\rangle$ with $|1,0\rangle$ and $|0,1\rangle$ for the Dresselhaus coupling can be written as
\begin{equation}
w_{0D1}=\frac{2\left(\Delta E\right)^3\left(eh_{14}\right)^2}{35\pi\hbar^4\rho}\left(\frac{1}{s_l^5}+\frac{4}{3s_t^5}\right)\left(|M_x|^2+|M_y|^2\right),
\label{W0D1}
\end{equation}
where $M_x = \langle \psi_i|x|\psi_f\rangle$ and $M_y=\langle \psi_i|y|\psi_f\rangle$. Beyond the dipole approximation, the spin-relaxation rate for the case of mixed Rashba and Dresselhaus spin-orbit coupling due to longitudinal and transverse phonons is written as
\begin{widetext}
\begin{eqnarray}
w_{0l}&=& \frac{\Delta E^7\left(eh_{14}\right)^2}{2\pi\rho\hbar^8 s_l^9} \frac{1}{9009}
\begin{bmatrix}
7|\langle \psi_i|x^3+y^3|\psi_f\rangle|^2+18Re \{\langle \psi_i|x^3|\psi_f\rangle\left(\langle \psi_i|xy^2|\psi_f\rangle\right)^\ast\} 
\\
+18Re \{\langle \psi_i|y^3|\psi_f\rangle \left(\langle \psi_i|x^2y|\psi_f\rangle\right)^\ast\}+27 |\langle \psi_i|x^2y+xy^2|\psi_f\rangle|^2
\end{bmatrix},
\label{W01}
~\nonumber\\
w_{0t}&=& \frac{\Delta E^7\left(eh_{14}\right)^2}{36\pi\rho\hbar^8 s_t^9} \frac{1}{9009}
\begin{bmatrix}
\frac{5}{429}|\langle \psi_i|x^3+y^3|\psi_f\rangle|^2+\frac{328}{15015}Re \{\langle \psi_i|x^3|\psi_f\rangle \left(\langle \psi_i|xy^2|\psi_f\rangle\right)^\ast\} \\
+\frac{328}{15015}Re\{\langle \psi_i|y^3|\psi_f\rangle \left(\langle \psi_i|x^2y|\psi_f\rangle\right)^\ast\}
+\frac{164}{5005} |\langle \psi_i|x^2y+xy^2|\psi_f\rangle|^2
\end{bmatrix}.
\label{W0t}
\end{eqnarray}
\end{widetext}
Due to the presence of two transverse phonon modes, we can write the total spin-relaxation rate for the mixed Rashba and Dresselhaus coupling case as $w_{0RD}= w_{0l}+2 w_{0t}$.

\begin{table}[b]
\caption{\label{table1}%
The material constants used in our calculations are taken from Ref.~\onlinecite{prabhakar13} unless otherwise stated.
}
\begin{ruledtabular}
\begin{tabular}{llcdr}
Parameters & GaAs & InAs & GaSb & InSb \\
\colrule
$g_{hh}$ &2.5$^a$  & -2.2$^a$ & -3.0^e & 3.0$^f$ \\
m & 0.14$^a$ & 0.115$^a$ & 0.35^f& 0.32$^g$\\
$\gamma_R~[{\mathrm{\AA}^2}]$ & 4.4 & 110 & 33& 500\\
$\gamma_D~[\mathrm{eV{\AA}^3}]$ & 26$^a$ &130$^a$ & 187 & 228\\
$\Delta~[\mathrm{meV}]$ & 346$^b$ &380$^b$ & 756^b&810$^b$\\
$\gamma_0$ & 6.85$^c$ &20$^d$ & 19.7^g&35.08$^h$\\
$E_g (\mathrm{eV})$ & 1.519$^b$ &0.417$^b$ & 0.822^b&0.235\\
$\eta=\Delta_{so}/\left(E_g+\Delta_{so}\right)$ &0.186$^e$& 0.477$^e$ & 0.48^e& 0.775$^e$\\
$eh_{14}~[10^{-5}\mathrm{erg/cm}]$ & 2.34 &0.54 & 1.5 & 0.75\\
$s_l~[10^{5}\mathrm{cm/s}]$ & 5.14 &4.2 & 4.3 & 3.69\\
$s_t~[10^{5}\mathrm{cm/s}]$ & 3.03 &2.35 & 2.49 & 2.29\\
$\rho~[\mathrm{g/cm^3}]$ & 5.3176 &5.667 & 5.6137 & 5.7747\\
\end{tabular}
\end{ruledtabular}
$^a$Reference~\cite{bulaev05},
$^b$Reference~\cite{gmitra16},
$^c$Reference\cite{hess76,shanabrook89},
$^d$Reference~\cite{vurgaftman01},
$^e$Reference~\cite{ganjipour15},
$^f$Reference~\cite{ghezzi95},
$^g$Reference~\cite{livneh12},
$^e$ The value of $\eta$ is  consistent with Ref.~\cite{bulaev05},
$^f$Reference~\cite{terent13}, $^g$Reference~\cite{pidgeon66},
$^h$Referebce~\cite{Boujdaria01}
\end{table}

\section{Computational Method}\label{computational-method}

We suppose that a  QD is formed at the center of a $1200\times 1200~\mathrm{nm^2}$ geometry.  We then diagonalize the total Hamiltonian $H$ numerically using the Finite Element Method~\cite{comsol}.  Since the geometry is much larger compared to the actual lateral size of the QD, we impose Dirichlet boundary conditions, find the  eigenvalues, eigenfunctions and the matrix elements $M\left(\mathbf{q}\alpha\right)$ of $H$.   The material constants for the simulations are taken from Table~\ref{table1}.
\begin{figure*}
\includegraphics[width=17cm,height=13cm]{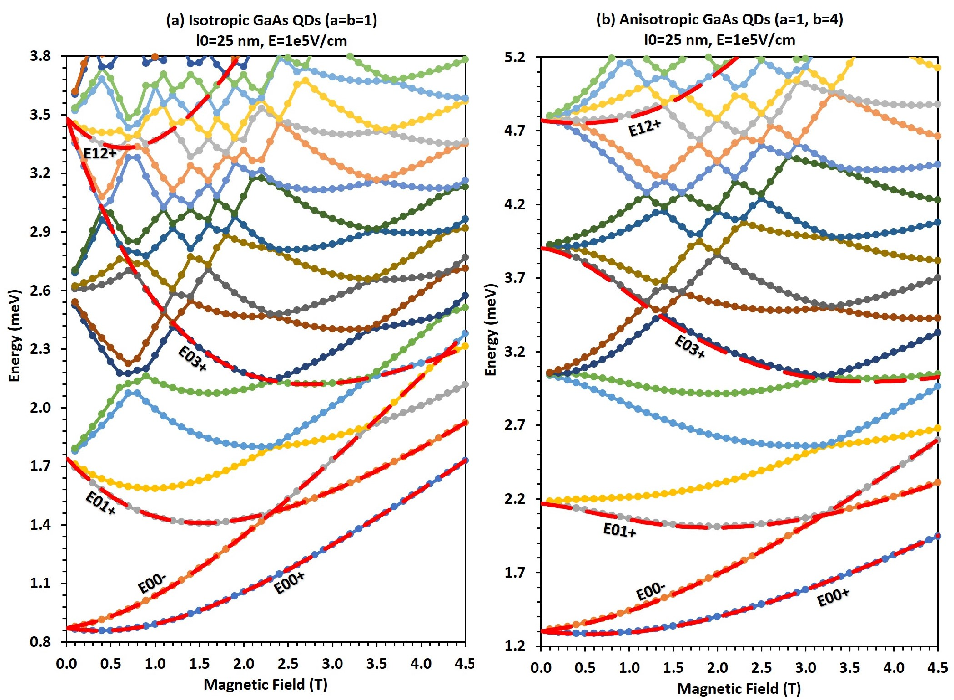}
\caption{\label{fig1GaAs} Energies of hole states in GaAs QDs are obtained analytically from Eq.~(\ref{epsilon-a}) (dashed lines, red color) and an exact diagonalization method (filled circles with solid lines).  As can be seen, the influence of spin-orbit couplings on the unperturbed eigenenergy is not profound, but wavefunctions are influenced in such a way that the interaction of phonons with heavy holes via spin-orbit coupling allows for the phonon mediated spin-relaxation time or decoherence time ($T_2 \approx 2T_1$) to be calculated. Note that the interaction of $\ket{01+}$, $\ket{03+}$ and $\ket{12+}$ states with $\ket{00-}$ states are of interest to calculate the phonon mediated spin-relaxation rate of hole states between $\ket{00-}$ and $\ket{00+}$ [see Eqs.~(\ref{HR-1}) and (\ref{HD-1})]. We chose an average dot height of 5 nm. }
\end{figure*}
\begin{figure*}
\includegraphics[width=17cm,height=5.8cm]{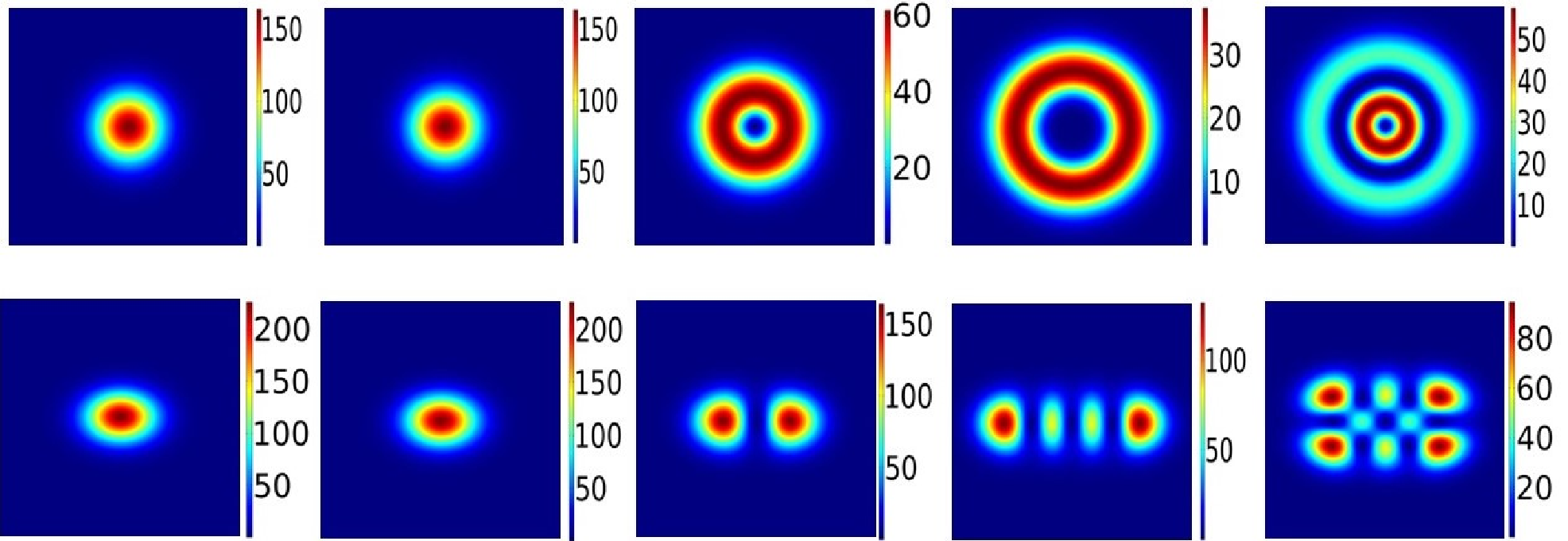}
\caption{\label{fig2GaAs} Probability density of valance band states of isotropic (a=b=1) (upper panel) and anisotropic (a=1, b=4) (lower panel) GaAs QDs. From left to right, the density of quantum dot states corresponds to $\ket{00+}$, $\ket{00-}$, $\ket{01+}$, $\ket{03+}$ and $\ket{12+}$. As can be seen from Eq.~(\ref{HD-1}), $\ket{01+}$ and $\ket{12+}$ states are responsible to flip the spin between $\ket{00-}$ and $\ket{00+}$ due to the Dresshelhaus spin-orbit coupling, while from Eq.~(\ref{HR-1}), $\ket{03u}$ is responsible for flipping the spin due to the Rashba spin-orbit coupling. We chose $\ell_0 = 25$ nm, $B=0.5$ T, $E = 10^5$ V/cm and average dot height = 5 nm. }
\end{figure*}

\section{Results and Discussions}\label{results-and-discussions}

The diagonalization of the Hamiltonian~(\ref{H}) finds a set of eigenvalues and eigenfunctions. As can be seen from  operators algebra results, Eqs.~(\ref{HR-1}) and~(\ref{HD-1}), the Rashba spin-orbit coupling is responsible for the admixtures of spin states $|0,0,-\rangle$ to $|0,3,+\rangle$ and the Dresselhaus spin-orbit coupling is responsible for the admixtures of spin states $|0,0,-\rangle$ to $|0,1,+\rangle$ and $|1,2,+\rangle$ for GaAs and InSb. This is because these materials possess a positive bulk g-factor. On the other hand, for InAs and GaSb, the intermixing of spin states $|0,0,+\rangle$ to $|0,3,-\rangle$ is due to the Rashba coupling, while the Dresselhaus coupling is accountable for the mixing between spin states $|0,0,+\rangle$ to $|0,1,-\rangle$ and $|1,2,-\rangle$. This effect is due to the fact that InAs and GaSb display negative bulk g-factor. In any case, these spin states provide the level crossing at several different values of the magnetic field and the QD radius. Hence, it is important to identify the exact and ideal location of the level crossing points so that the spin hot spots can be recognized in the phonon mediated spin-relaxation rate of III-V semiconductor QDs.  In Fig.~\ref{fig1GaAs}, we have plotted the band structures of GaAs QDs obtained from the exact diagonalization technique (solid circles) and compared them with the analytical results obtained from Eq.~(\ref{epsilon-a}) for both symmetric QDs (a=b=1) (dashed red lines) in Fig.~(\ref{fig1GaAs}(a)) and asymmetric QDs (a=1,b=4) (dashed red lines) in Fig.~(\ref{fig1GaAs}(b)). We have only shown the analytical results (dashed red lines) involving some particular states of interest that interact with the lowest spin up and down states mediated by phonon. As can be seen, the analytically obtained band structures from unperturbed Hamiltonian are in good agreement with the band structures obtained from the exact diagonalization method because the spin-orbit coupling being weak has almost no influence on the eigenvalues of the bands. However, identifying states $|0,0,+\rangle$, $|0,0,-\rangle$, $|0,1,+\rangle$, $|0,3,+\rangle$, and $|1,2,+\rangle$ in the bands of GaAs QDs obtained from exact diagonalization method, is beneficial when we calculate the matrix elements of the spin-relaxation rate between several different states of the dots (see Eqs.~\ref{W0D1},\ref{W01},\ref{W0t}).

\begin{figure*}
\includegraphics[width=18cm,height=8.0cm]{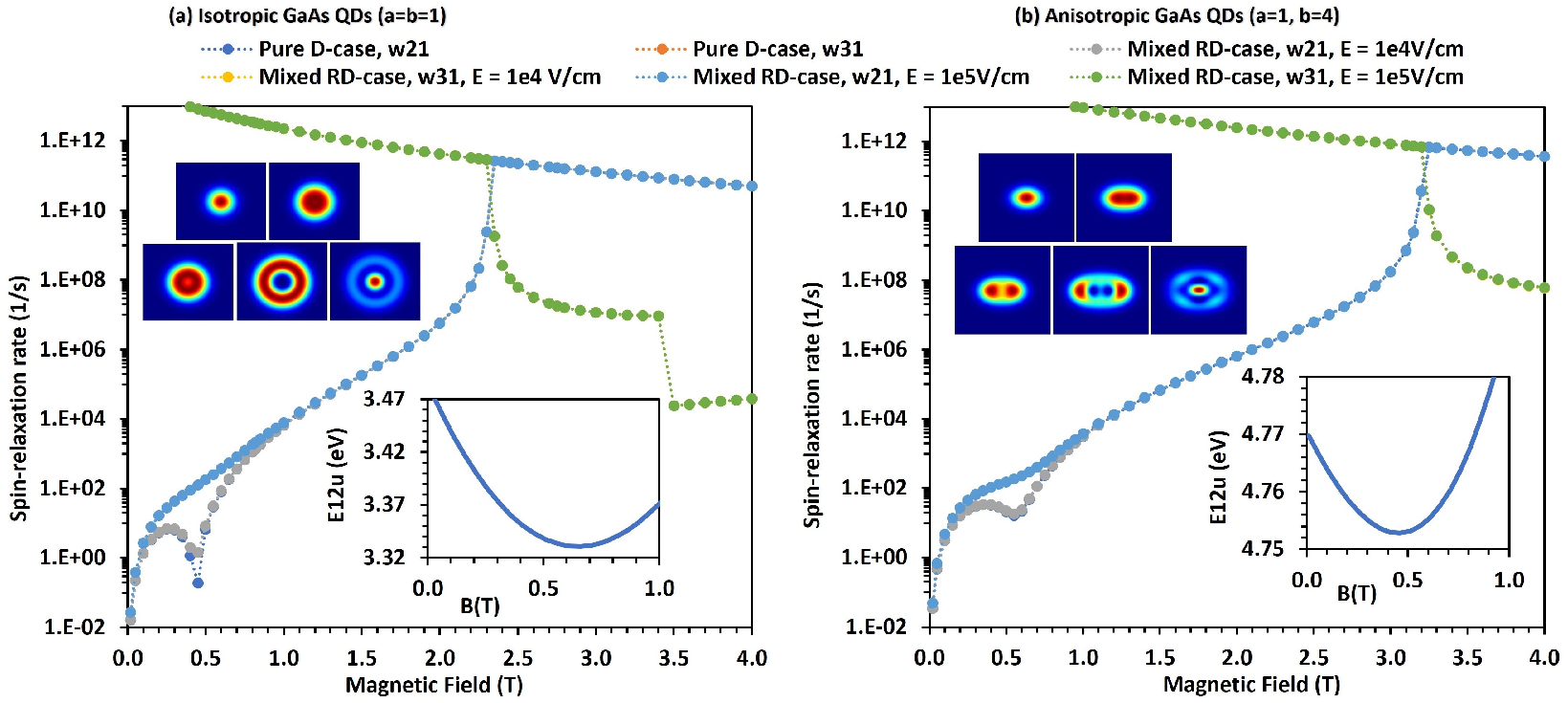}
\caption{\label{fig3GaAs} (Color online) Spin-relaxation rate of the heavy hole versus magnetic field in isotropic (a) and anisotropic (b) GaAs QDs. The spin hot spot, i.e.\ the cusp-like structure, can only be observed for the pure Dresselhaus spin-orbit coupling case due to the strong admixture of higher states to the heavy-hole spin states. The spin-relaxation rate is seen as a monotonous function of the magnetic field for the pure Rashba coupling case (see Fig.~\ref{figA} in Appendix). Also, at low magnetic fields ($B<0.35$ T), nonlinearity in the spin-relaxation rate is found for pure Dresselhaus case. This nonlinearity is due to the band curvature of the state $|1,2,+\rangle$ (see insets). The spin hot spot is always present in the anisotropic QDs due to the broken in-plane rotational symmetry. Energies of hole states, as in Fig.~\ref{fig1GaAs}, are shown for isotropic and anisotropic dots. Probability densities of the states where the spin hot spot is observed, as in Fig.~\ref{fig2GaAs}, are presented in the other insets in (a) and (b). Here, w21 represents the spin-relaxation rate between the states $|0,0,-\rangle$ and $|0,0,+\rangle$ and w31 represents the spin-relaxation rate between $|0,1\rangle$ and $|0,0\rangle$ that have same spin states. The letters R, D and RD correspond to the pure Rashba, pure Dresselhaus and mixed Rashba-Dresselhaus couplings. We chose $\ell_0 = 25$ nm and average dot height = 5 nm. }
\end{figure*}

\begin{figure*}
\includegraphics[width=18cm,height=19cm]{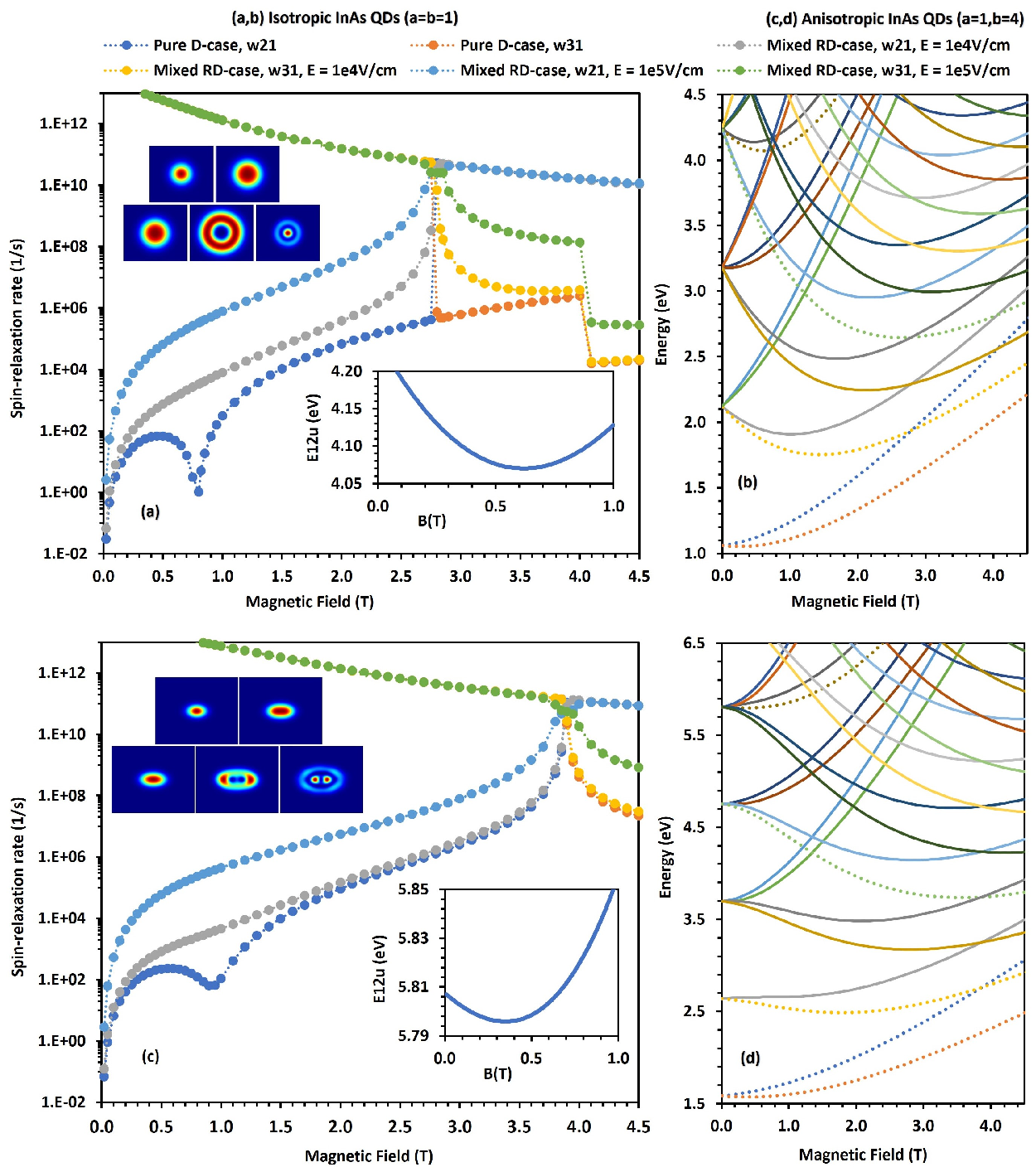}
\caption{\label{fig4InAs} Same as in Fig.~\ref{fig3GaAs} but for InAs heavy-hole QDs. For the isotropic QDs (a), the spin hot spot is seen for the Rashba (R) coupling due to the strong admixture of higher states to the heavy-hole spin states (see Fig.~\ref{figA} in Appendix). The spin-relaxation rate is a monotonous function of the magnetic field for the Dresselhaus (D) coupling for isotropic QDs. Spin hot spot is always present in the anisotropic QD due to the broken in-plane rotational symmetry. Also, at low magnetic fields ($B<1$ T), nonlinearity in the spin-relaxation rate is seen for the Dresselhaus coupling (see insets for band curvatures). Isotropic and anisotropic probability densities (insets) of spin states at magnetic fields, where spin hot spot is present, are shown. The energy of isotropic and anisotropic hole states are respectively in panel (b) and (d).  For mixed RD-case at large electric fields, the spin hot spot is shown to cover a wide range of magnetic fields. Here, w21 represents the spin-relaxation rate between the states $|0,0,+\rangle$ and $|0,0,-\rangle$ and w31 represents the spin-relaxation rate between $|0,1\rangle$ and $|0,0\rangle$ that have same spin states. We chose $\ell_0=25$ nm and average dot height = 5 nm. }
\end{figure*}

\begin{figure*}
\includegraphics[width=18cm,height=19cm]{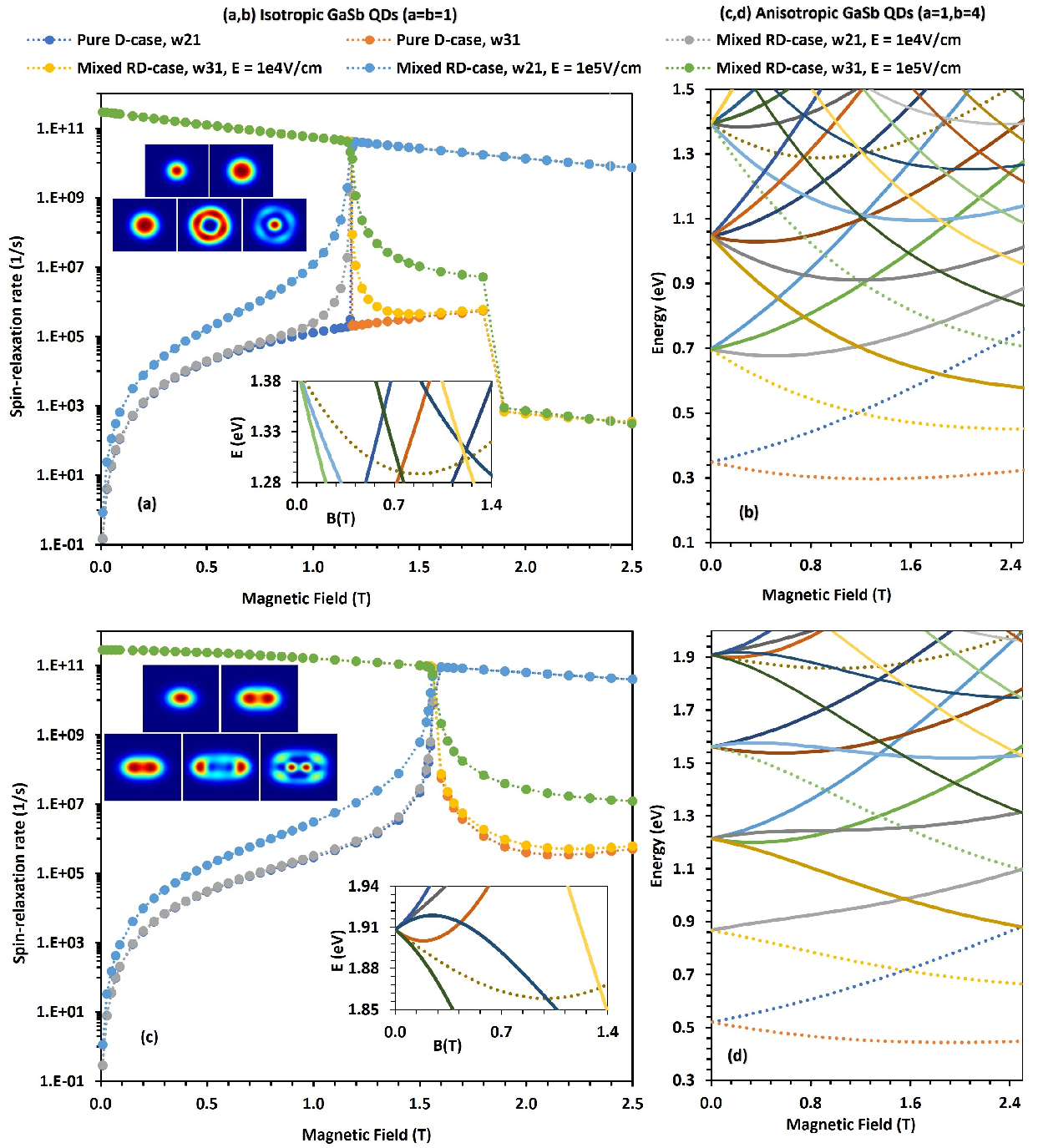}
\caption{\label{fig5-GaSb} Same as in Fig.~\ref{fig3GaAs} but for GaSb heavy-hole QDs. For the isotropic QD (a), the spin hot spot is seen for the Rashba (R) coupling due to the strong admixture of higher states to the heavy-hole spin states (see Fig.~\ref{figA} of Appendix). The spin-relaxation rate is a monotonous function of magnetic field for the Dresselhaus (D) coupling. Spin hot spot is always present in the anisotropic QD (c) due to the broken in-plane rotational symmetry. Also, at low magnetic fields, we can not find nonlinearity in the spin-relaxation rate for the Dresselhaus case due to the level crossing (see insets). Isotropic and anisotropic
probability densities (insets) of spin states at magnetic fields, where spin-hot spot is present, are shown. The energy of isotropic and anisotropic hole states are respectively in panels (b) and (d). For mixed RD-case at large electric fields, the spin hot spot is shown to cover a wide range of magnetic fields. Here, w21 represents the spin-relaxation rate between the states $|0,0,+\rangle$ and $|0,0,-\rangle$ and w31 represents the spin-relaxation rate between the states $|0,1\rangle$ and $|0,0\rangle$ that have same spin states. We chose $\ell_0=25$ nm and  $E =10^5$ V/cm. }
\end{figure*}

\begin{figure*}
\includegraphics[width=18cm,height=19cm]{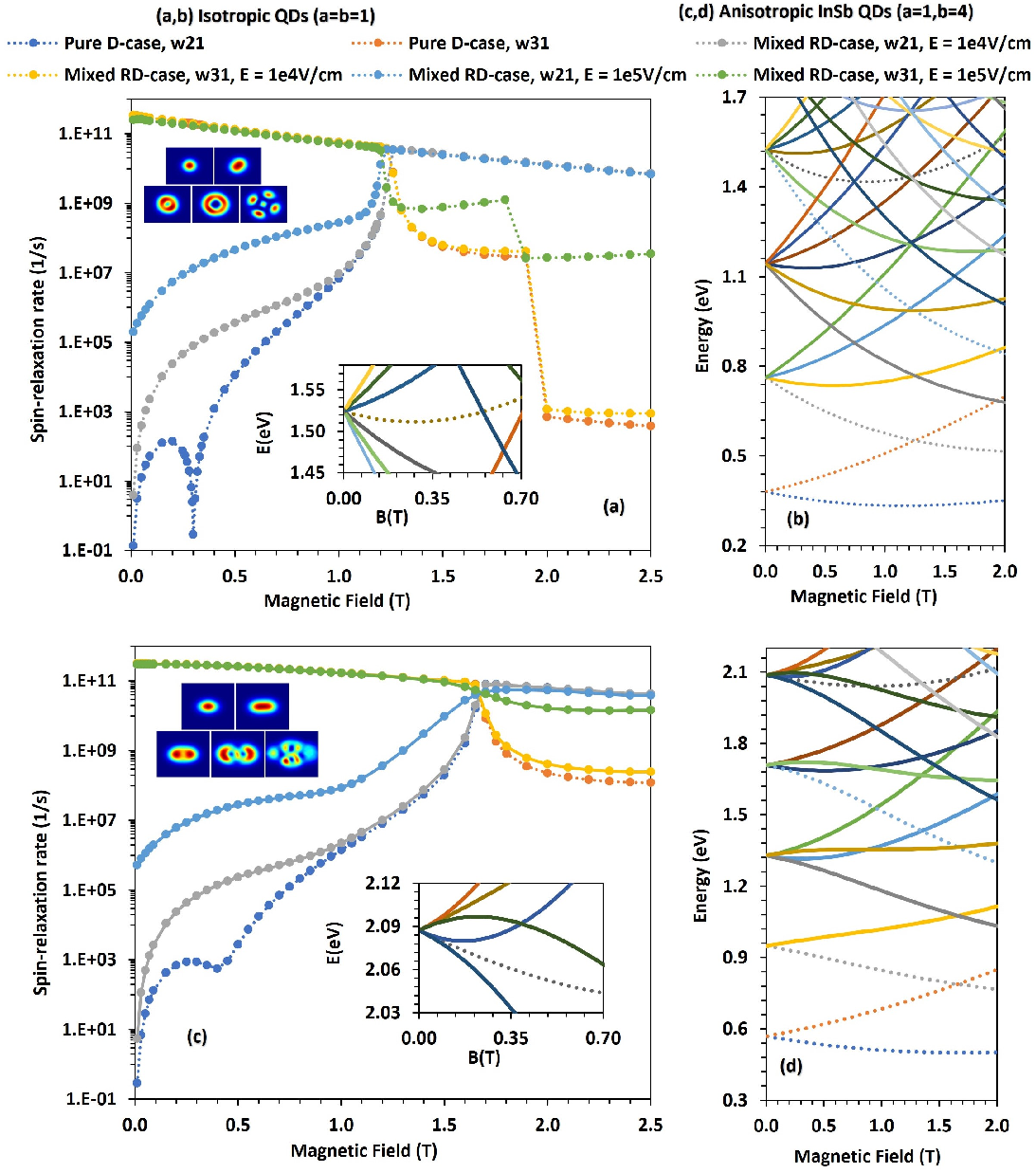}
\caption{\label{fig6-InSb} Same as in Fig.~\ref{fig3GaAs} but for InSb heavy-hole QDs. For the isotropic QD (a), the spin hot spot is seen for the Dresselhaus (D) coupling due to the strong admixture of higher states to the heavy-hole spin states. The spin-relaxation rate is a monotonous function of the magnetic field for the Rashba (R) coupling (see Fig.~\ref{figA} of Appendix). Spin hot spot is always present in anisotropic QD (c) due to the broken in-plane rotational symmetry. Also, at low magnetic fields ($B<0.3$ T) for the isotropic and ($B< 0.5$ T) for the anisotropic QD, nonlinearity in the spin-relaxation rate is observed for the Dresselhaus case (see inset plots of $E$-vs-$B$ in Fig.~\ref{fig6-InSb} (a) and (c)). Isotropic and anisotropic probability densities (insets) of spin states at magnetic fields, where spin-hot spot is present, are shown. The energy of isotropic and anisotropic hole states are respectively in panels (b) and (d).  For mixed RD-case at large electric fields, the spin hot spot is shown to cover a wide range of magnetic fields.  Here, w21 represents the spin-relaxation rate between the states $|0,0,+\rangle$ and $|0,0,-\rangle$ and w31 represents the spin-relaxation rate between the states $|0,1\rangle$ and $|0,0\rangle$ that have same spin states. We chose $\ell_0=25$ nm and an average dot height of 5 nm. }
\end{figure*}

\begin{figure*}
\includegraphics[width=17cm,height=14cm]{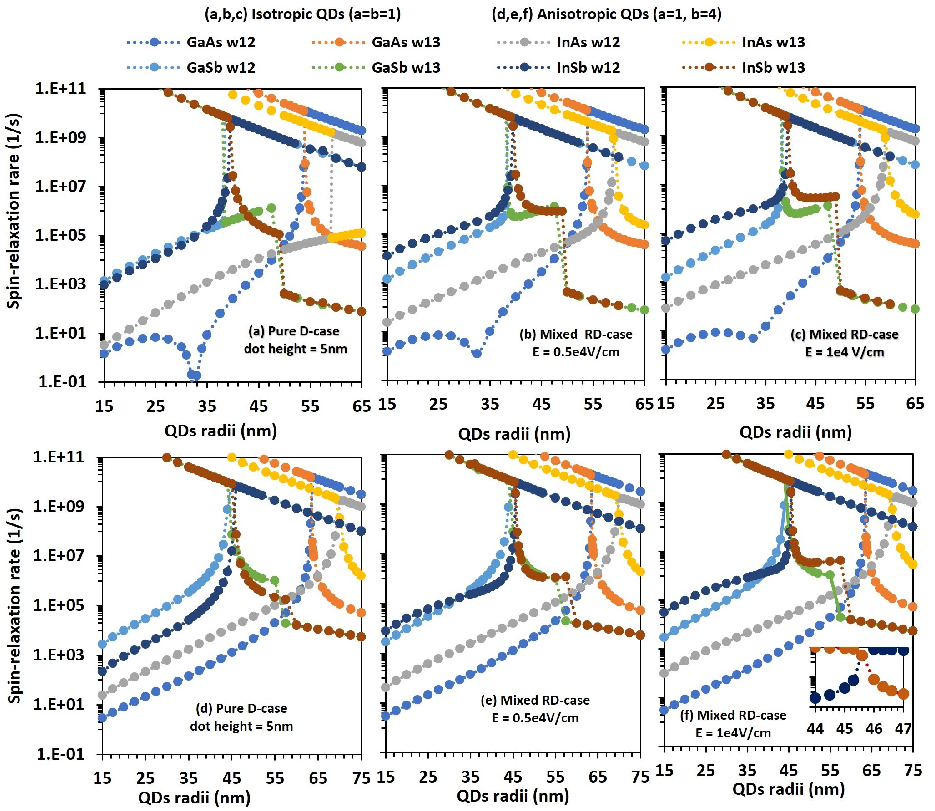}
\caption{\label{fig7} Phonon mediated spin-relaxation rate vs QD radii of isotropic (upper panel) and anisotropic (lower panel) GaAs, GaSb, InAs, and InSb for the Dresselhaus spin-orbit coupling (a,d) and mixed RD-cases (b,c,e,f). For isotropic QDs, the spin hot spots are present in InAs and GaSb QDs for the pure Rashba spin-orbit coupling (see Fig.~\ref{figA} in Appendix). But the spin hot spot is present in GaAs and InSb QDs for the pure Dresselhaus spin-orbit coupling (a). Note that GaAs and InSb have positive bulk g-factor, whereas InAs and GaSb have negative bulk g-factor of heavy holes. Also in (a) for GaAs at around 34-nm radius, we observe a minimum in the spin-relaxation rate due to the interaction of the phonon with heavy-hole spin-orbit couplings between the states $|1, 2, +\rangle$ and $|0, 0,-\rangle$. In anisotropic QDs in (d), (e) and (f), the spin hot spot is always present due to the broken in-plane rotational symmetry. Here, w21 represents the spin-relaxation rate between the states $|0,0,+\rangle$ and $|0,0,-\rangle$ and w31 represents the spin-relaxation rate between the states $|0,1\rangle$ and $|0,0\rangle$ that have same spin states. We chose $B=0.5$ T. }
\end{figure*}

In Fig.~\ref{fig2GaAs}, we plot the probability density of the states $|0,0,+\rangle$, $|0,0,-\rangle$, $|0,1,+\rangle$, $|0,3,+\rangle$ and $|1,2,+\rangle$ for isotropic QDs (a=b=1) (upper panel from left to right) and for anisotropic QDs (a=1,b=4) (lower panel from left to right). Note that these states of heavy hole in QDs can only interact with phonons within the dipole approximation for the Dressehaus spin-orbit coupling but for both the Rashba and Dresselhaus spin-orbit couplings beyond the dipole approximation.   As can be seen in these figures, the anisotropy has a significant influence on the probability density of the heavy-hole wave function which induces the spin hot spot regardless of the pure Rashba or the pure Dresselhaus couplings.  This effect is a direct consequence of the broken in-plane rotational symmetry. For the heavy hole in QDs, it is therefore expected that the anisotropy ($a\neq b$ in Eq.~(\ref{H0})) will always induce spin hot spots both for the cases of pure Rashba and pure Dresselhaus couplings. Our results in Figs.~\ref{fig3GaAs}, \ref{fig4InAs}, \ref{fig5-GaSb}, \ref{fig6-InSb}, and \ref{fig7} in general predict this universal trend. A similar anisotropic effect for the electron spin hot spot has also been observed in the cases of Rashba and Dresselhaus spin-orbit couplings~\cite{prabhakar13}. On the other hand, we find that for heavy holes in isotropic QDs, the spin hot spot for the pure Rashba and the pure Dresselhaus couplings critically depends on the bulk g-factor of the heavy hole (also see Figs.~\ref{fig3GaAs}, \ref{fig4InAs}, \ref{fig5-GaSb}, \ref{fig6-InSb}, \ref{fig7} and \ref{figA}).

In Fig.~\ref{fig3GaAs}(a) for the isotropic GaAs QD, we observe that the spin-relaxation rate is a monotonous function of the magnetic field for the pure Rashba coupling (see Fig.\,\ref{figA}(a) of  Appendix), while the spin hot spot, i.e. the cusp-like structure, is present for the pure Dresselhaus coupling. However, in Fig.~\ref{fig3GaAs}(b) for the anisotropic GaAs QD, we find that the spin hot spot is present for both the pure Rashba  and the pure Dresselhaus cases. This is a direct consequence of the broken in-plane rotational symmetry caused by the anisotropy. In Ref.~\cite{prabhakar13} for electrons in an isotropic and anisotropic QD, we have previously shown that the spin hot spots originate due to the strong interaction of the spin states with other available higher states mediated by phonon~\cite{takahashi10}. In fact, as can be seen in Eqs. (26) and (27) of Ref.~\cite{prabhakar13}, the spin hot spot will only be produced if there is a degeneracy in the denominator of the matrix element. Since we also observe the spin hot spot for heavy holes in both isotropic and anisotropic dots, it is reasonable to assume that such degeneracy will appear in the denominator of the matrix elements in the full calculation involving heavy holes within the framework of second order perturbation theory. The probability density of degenerate states (the last figure of upper panel and the first figure of lower panel) are shown within the inset panels of Fig~\ref{fig3GaAs}.  To be more precise, the matrix element of Eq.~(\ref{1-T1}) can be found as
\begin{widetext}
\begin{equation}\label{Mqa}
 M(q\alpha) = \sum_{n_+,n_-} \frac{ \langle 0,0  \arrowvert U_{ph} \arrowvert   n+,n_- \rangle  \langle n_+,n_-,-  \arrowvert H_{so} \arrowvert   0,0,+ \rangle  }{\varepsilon_{0,0,+}^0-\varepsilon_{n_+,n_-,+}^0}
 + \sum_{n_+,n_-} \frac{ \langle 0,0,-  \arrowvert H_{so} \arrowvert   n+,n_-,+ \rangle  \langle n_+,n_-  \arrowvert U_{ph} \arrowvert   0,0 \rangle  }{\varepsilon_{0,0,-}^0-\varepsilon_{n_+,n_-,+}^0},
\end{equation}
\end{widetext}
and $H_{so}$ can be written in terms of raising and lowering operators, as given in Eqs.~(\ref{HR-1}) and \ref{HD-1}. However, the calculation of the matrix element, $M(q\alpha)$, of Eq.~(\ref{Mqa}) for the heavy hole is rather cumbersome.

At low magnetic fields in the range of $0$T to $0.5$T for the case of pure Dresselhaus spin-orbit coupling in Figs.~\ref{fig3GaAs}(a) and (b), we find maxima in spin-relaxation rate due to the presence of minima in the bandstructures of $|1,2,+\rangle$ state; these minima are presented in the inset plots. Note that the band structures of $|1,2,+\rangle$ state interact with phonon through the spin-orbit coupling for the pure Dresselhaus case only (see Eq.~(\ref{HD-1})). To capture the resulting maximum points in the spin-relaxation rate, we expand the matrix element beyond the dipole approximation (see the fourth term on right hand side of Eq.~\ref{Mq}) so that the non-vanishing matrix element from the hole-phonon interaction between the states $|1,2,+\rangle$  and $|0,0,-\rangle$ can be calculated. There is a systematic appearance of maxima due to this nonlinearity in materials, such as GaAs, InAs, GaSb, at low magnetic fields. This is also true for small lateral sized QDs  of the heavy hole for the case of Dresselhaus spin-orbit coupling.  At around $3.5$T in Fig.\ref{fig3GaAs}(a), there is an abrupt change in the spin-relaxation rate due to the level crossing of $|0,0,-\rangle$ state to the other higher state (also see Fig.~\ref{fig1GaAs}(a)). In Fig.~\ref{fig3GaAs}, we also plot the spin-relaxation rate for mixed RD-cases at $E=10^4$V/cm and $E=10^5$V/cm. We notice that at $E=10^4$V/cm, the relaxation rate for mixed RD-case and pure Dresselhaus coupling are the same because the Rashba coupling is weaker than the Dresselhaus coupling. At $E=10^5$V/cm for mixed RD-case, enhancement in the spin-relaxation rate can be observed.

We plot the spin-relaxation rate and band structures of an isotropic InAs QD with respect to the magnetic field in, respectively, Fig.~\ref{fig4InAs}(a) and (b) (upper panels). Similarly, in Fig.~\ref{fig4InAs}(c) and (d) (lower panels), we plot the spin-relaxation rate and band structures of an anisotropic InAs QD as a function of the magnetic field. In contrary to the isotropic GaAs QD in Fig.~\ref{fig3GaAs} (i.e., where the spin hot spot is present for the pure Dresselhaus case but absent for the pure Rashba case), we find the spin hot spot for the pure Rashba case (see Fig.~\ref{figA} (a,b) in Appendix) but not for the pure Dresselhaus (D) case in InAs. This is due the fact that the bulk g-factor of the heavy-hole InAs is negative, whereas it is positive for GaAs (see Table~\ref{table1}). It is true that InAs is a small band gap material and has a large Rashba (R) spin-orbit coupling ($\alpha_{\mbox R}$ which is about one order magnitude larger than $\alpha_{\mbox D}$ at $E=10^5$V/cm). Yet, most importantly, its negative bulk g-factor of heavy holes induces different admixture mechanism in the interaction of the heavy hole with phonon. As can be seen by the dotted lines in Fig.~\ref{fig4InAs}(b) and (d)) for InAs QDs, the band structures of the $|0,0,+\rangle$ state interact with those of the $|0,1,-\rangle$ and $|1,2,-\rangle$ for the pure Dresselhaus case and with the $|0,3,-\rangle$ state for the pure Rashba (R) case. Notice that in comparison to the isotropic GaAs QD, a different admixture mechanism of heavy-hole phonon interaction in the isotropic InAs QD leads to the formation of the spin hot spot for the pure Rashba case, while, in contrast, it is absent for the pure Dresselhaus case. In an anisotropic InAs QD with the violation of the in-plane rotational symmetry (insets), the universality returns as in GaAs, and we always find a spin hot spot (see Eq.~(\ref{Mqa}).

In Fig.~\ref{fig4InAs}, we further plot the spin-relaxation rate for mixed RD-case at $E=10^4$V/cm and $E=10^5$V/cm. Here we find that spin-relaxation rate increases as we increase the electric fields. But most importantly, spin hot spot widens with an increase in electric fields. This confirms that the Rashba spin-orbit coupling covers a wide range of spin hot spot in any particular height of QDs. In this paper, we chose the average height of the dot to be 5 nm for all the materials considered (GaAs, InAs, GaSb, InSb).  In the inset of Fig.~\ref{fig4InAs}, we have plotted the probability density of the states at the spin hot spot associated with the bands shown by the dotted lines. The probability density of degenerate states (the last figure of the upper inset-panel and the first figure of the lower inset-panel) are clearly visible that induce the spin-hot spot in the dots. 

In Fig.~\ref{fig5-GaSb}, we plot the spin-relaxation rate and band structures versus the magnetic field for isotropic and anisotropic GaSb QDs that has a negative bulk g-factor. Similarly, in Fig.~\ref{fig6-InSb}, we present the spin-relaxation rate and band structures as a function of the magnetic field for isotropic and anisotropic InSb QDs that has a positive bulk g-factor. For the isotropic GaSb QD in Fig.~\ref{fig5-GaSb}(a) (also see Fig.\ref{figA} in Appendix), we notice that the spin hot spot is present for the Rashba case of heavy holes because, as noted above, of its negative bulk g-factor. For the isotropic InSb QD in Fig.~\ref{fig6-InSb}(a), conversely, the spin hot spot appears for the case of the pure Dresselhaus coupling because it has a positive bulk g-factor. For the anisotropic QD in Figs.~\ref{fig5-GaSb}(c) and \ref{fig6-InSb}(d), the heavy-hole spin hot spot can be observed for both Rashba  and Dresselhaus cases since the anisotropy disrupts the in-plane rotational symmetry. The insets of upper panel of Fig.~\ref{fig5-GaSb} and \ref{fig6-InSb} are the probability density of hole states at the spin hot spot for mixed RD-case. We again can clearly visualize the probability density of degenerate states
(the last figure of the upper-insets and the first figure of the lower-insets) that induce the spin hot spot in the dots. At low magnetic fields for the pure D-case of GaSb dots in Fig.~\ref{fig5-GaSb}, we do not find any general trend of maxima in spin-relaxation rate because there is a level crossing of the bands before the minima appears in the band structures of the $|1,2,u\rangle$ state (see inset plot of Fig.~\ref{fig5-GaSb}).

In Fig.~\ref{fig3GaAs}, the cusp-like structures near the spin-hot spot for mixed-RD cases at $E=10^4$V/cm and $E=10^5$V/cm are the same because GaAs possesses a weak Rashba spin-orbit coupling. However, in Figs.~\ref{fig4InAs}, \ref{fig5-GaSb}, and \ref{fig6-InSb}, we clearly observe that electric fields enhance the spin-relaxation rate. In fact, most importantly, the increasing electric field widens the cusp-like structures. Hence, we conclude that the electric fields couple to an wide range of magnetic fields at the spin-hot spot.

In Fig.~\ref{fig7}, we display the phonon mediated spin-relaxation rate versus the dot radius for both the isotropic (upper panel) and anisotropic (lower panel) geometries of GaAs, GaSb, InAs, and InSb QDs. In the context of isotropic QDs in Fig.~\ref{fig7}(a) for the pure Dresselhaus case, we again find that the spin hot spots, i.e. the cusp-shaped structures, are present in both GaAs and InSb. This is because, as again, GaAs and InSb characterize positive bulk g-factors of the heavy hole. On the other hand, the spin hot spot is absent in InAs and GaSb QDs owing to their negative bulk g-factors for the heavy hole. Furthermore, in Fig.~\ref{fig7}(a) for a GaAs QD at low magnetic fields, the minima in the phonon mediated spin-relaxation rate can be seen as due to the interaction of the phonon with heavy-hole spin-orbit couplings between the states $|1,2,+\rangle$ and $|0,0,-\rangle$. In Fig.~\ref{fig7}(b) and (c), we plot the spin-relaxation rate versus the dot radius of GaAs, GaSb, InAs and InSb materials for mixed RD cases at $E=5\times 10^3$V/cm and  $E=1\times 10^4$V/cm. Here, in addition to the enhancement of spin-relaxation rate, the spin hot spot always exists for mixed RD-cases. For isotropic QDs, we therefore conclude that if the spin hot spot is absent for the pure Dresselhaus case (InAs and GaSb dots in Fig.\ref{fig7}(a)), then it must be present due to the Rashba spin-orbit coupling (InAs and GaSb dots of Fig.~\ref{figA} in Appendix). Reciprocally, in case the spin hot spot is present for the pure Dresselhaus case (GaAs and InSb dots in Fig.\ref{fig7}(a)), then it must be forbidden within in the Rashba coupling (see GaAs and InSb dots in Fig.~\ref{figA} in Appendix). In Fig.~\ref{fig7}(d,e,f) for anisotropic QDs, the spin hot spot can be seen in both the Dresselhaus and the Rashba coupling (see Fig.~\ref{figA} in Appendix) scenarios owing to the broken in-plane rotational symmetry. Notice that a very sharp spin hot spot shows up for the InSb QD for the mixed RD coupling (see the inset plot of Fig.~\ref{fig7}(f)) because it possesses a large Rashba coupling coefficient (see Table \ref{table1}). Thus, using the size analysis, we re-confirm that in the interaction of a heavy hole with the phonon, the strong intermixing of heavy-hole spins eventually induces spin hot spots, and is controlled by the effective bulk g-factor of the heavy hole.

In Ref.~\cite{amasha08}, an experimental study of spin-relaxation rate in quantum dots due to the piezophonon finds that the rate varies up to $\approx 10^3s^{-1}$. In the current study, we find that heavy-hole spin-relaxation rate is comparable to the experiment~\cite{amasha08} at low magnetic fields and small dot sizes, while the rate is several orders of magnitude larger than the experiment in a regime where the spin hot spot is observed. Measurement of such a fast spin-relaxation rate comes from another channel -- the direct spin-phonon coupling between the Zeeman sublevels of the orbital state due to acoustic phonons~\cite{khaetskii01,frenkel91}.  Below, we present results for heavy-hole spin-relaxation rate due to emission or absorption of a single piezophonon where experiments can likely be conducted~\cite{amasha08}.
\begin{figure*}
\includegraphics[width=18cm,height=7cm]{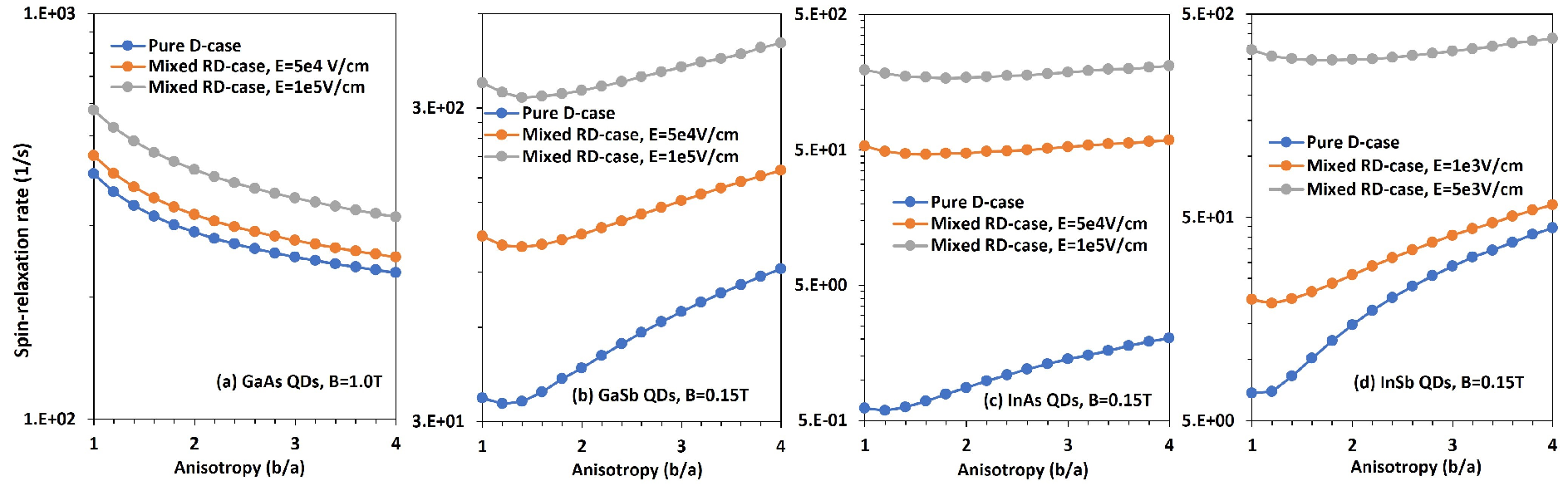}
\caption{\label{fig8} Phonon mediated spin-relaxation rate vs the anisotropy in QDs. Enhancement in spin-relaxation rate with electric field can be seen when both the Rashba and the Dresselhaus couplings are present.  We chose the parameters in such a way that spin-relaxation rate lies up to $10^3 s^{-1}$ so one can perform experiment to detect spin-relaxation rate using an experimental setup similar to Ref.\cite{amasha08}. We chose $\ell_0=15$ nm. }
\end{figure*}

In Fig.~\ref{fig8}, we exhibit the variation of the spin-relaxation rate versus the anisotropy of GaAs, GaSb, InAs and InSb QDs respectively on panels (a,b,c, and d) for the pure Dresselhaus and mixed Rashba-Dresselhaus spin-orbit coupling cases. We chose all the control parameters, magnetic fields, electric fields and size of the dots, in such a fashion that the spin-relaxation rate for the materials lies within the rate observed in the experiment~\cite{amasha08}. Enhancements in the relaxation rate can be seen for the mixed Rashba-Dresselhaus coupling case. Since the Rashba coupling is stronger in InAs and InSb QDs, the influence of electric fields on spin-relaxation rates for these materials is significantly larger than GaAs and GaSb dots.

\begin{figure}
\includegraphics[width=8.5cm,height=7cm]{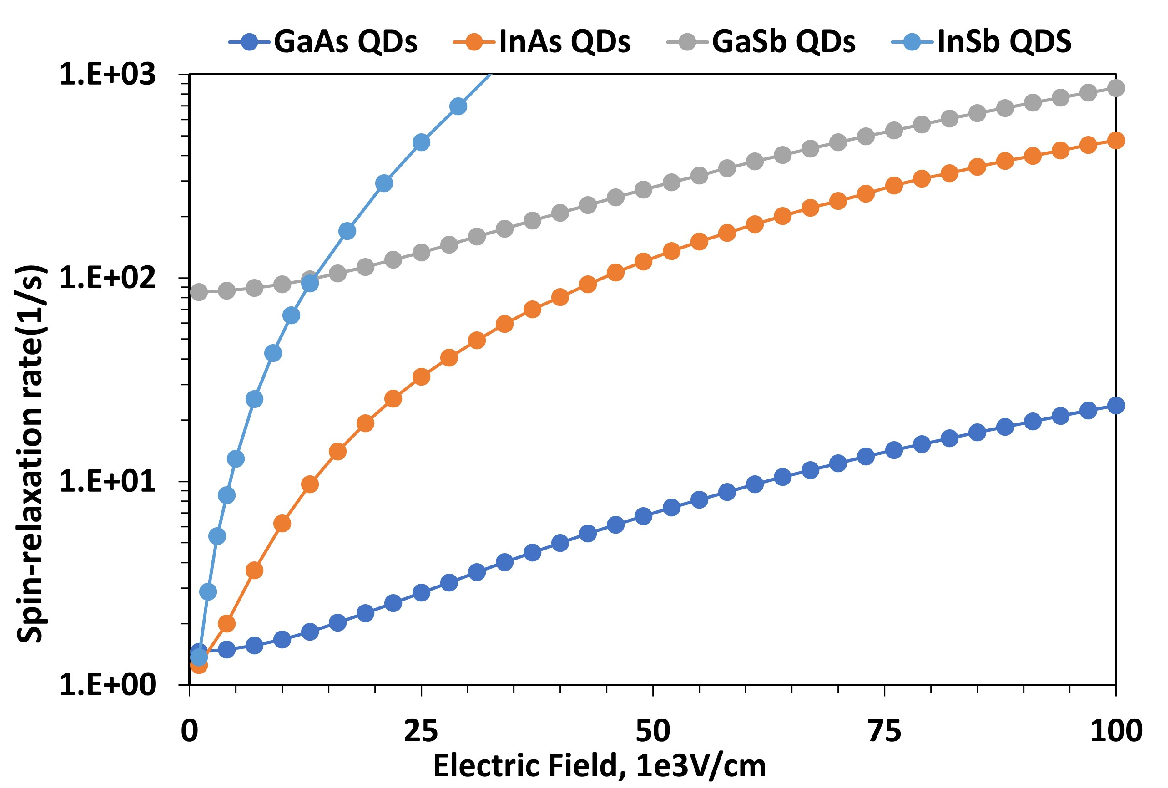}
\caption{\label{fig9} Phonon mediated spin-relaxation rate vs electric field in isotropic QDs (a=b=1). We chose the parameters (B=0.5 T for GaAs, 0.2 T for InAs and GaSb and 0.05 T for InSb) in such a way that the spin-relaxation rate lies up to $10^3 s^{-1}$ and one can perform experiment to detect the spin-relaxation rate using the experimental setups as in Ref.\cite{amasha08}. We chose $\ell_0=15$. }
\end{figure}

In Fig.~\ref{fig9}, we delineate the spin-relaxation rate with the variation of the electric field in isotropic GaAs, InAs, GaSb and InSb dots. It is clear that as we increase the electric field, spin-relaxation rate increases due to the contribution coming from the Rashba spin-orbit coupling. In InSb dots, there is a sharp enhancement in the relaxation rate because InSb possesses large Rashba spin-orbit coupling strength. We may notice that the control parameters in Figs.~\ref{fig8} and \ref{fig9} are chosen in such a way that the spin-relaxation rate still varies up to $10^3s^{-1}$ such that one may ensure that such spin states can be accessed experimentally by low temperature measurements~\cite{amasha08}.

The calculated spin-relaxation rate in Figs.~\ref{fig8} and \ref{fig9} is small and lies within the range of experimental values for electrons at low temperature measurements~\cite{amasha08}. This suggests, our theoretical model of heavy-hole interaction with emission or absorption of single piezophonon works well because we consider a small
phonon density of states at the scale of the Zeeman energy. On the other hand, for the spin-relaxation rate at the spin hot spots, where the rates are several order of magnitude larger than the experimental values~\cite{amasha08}, the mechanism of spin flipping from other channels (e.g. due to acoustic phonon, multi-phonon processes, modulation of the hyperfine coupling with nuclei by lattice vibrations, exchange scattering process) may become relatively more important than the single-piezophonon processes~\cite{frenkel91,gantmakher-book,khaetskii00,khaetskii01}. This is because with the rising temperature the probability of absorption and emission of phonon is increased.  The derivation of spin-relaxation rate for two-phonon processes and other mechanisms within the channel  is explicitly presented in Ref.~\cite{khaetskii01} and one may improve our model to capture the influence of such mechanisms on the phonon mediated transition rate of heavy hole in QDs.

\section{Conclusions}\label{conclusion}

In this study, we have demonstrated that the phonon mediated spin-relaxation rate of heavy holes in III-V semiconductor QDs can be controlled with electric fields, magnetic fields and the lateral size of the dots. The characteristic spin hot spot, which is found in the spin-relaxation  rate, shows a striking dependence on the bulk g-factor of the heavy hole. Our calculations show an enhancement of the heavy-hole spin-relaxation rate in narrow band gap materials, namely InAs, InSb and GaSb, due to their strong Rashba spin-orbit coupling interactions. In isotropic GaAs and InSb QDs in Figs.\,\ref{fig3GaAs}(a), \ref{fig6-InSb}(a) and \ref{fig7}(a), we have shown that the spin hot spot, i.e., the cusp-like structure, can be induced in the phonon mediated spin-relaxation rate for the pure Dresselhaus spin-orbit couplings. On the other hand, for these materials, the spin hot spot is found absent, i.e. the spin-relaxation rate is a monotonous function of the magnetic field and QD-radius for the pure Rashba coupling case (Fig.~\ref{figA} in Appendix). Both these effects are due to the fact that these materials possess positive bulk g-factor of heavy holes.  However, in isotropic InAs and GaSb QDs in Fig.\,\ref{figA} in Appendix, our results have demonstrated that the spin hot spot can be seen in phonon mediated spin-relaxation rate for the pure Rashba couplings, whereas it is absent for the pure Dresselhaus case~(Figs.\,\ref{fig4InAs}, \ref{fig5-GaSb}). This effect owes to the fact that these materials possess negative bulk g-factor of heavy holes. Hence, we conclude that the detection of spin hot spot is very sensitive to the bulk g-factor of heavy holes in QDs and therefore to the specific choice of the QD material. For anisotropic QDs, on the other hand, the spin hot spot is universally present for both the Rashba and the Dresselhaus spin-orbit coupling cases due to the broken in-plane rotational symmetry effect induced by the anisotropy. This is evidenced in the probability density distribution of the valence band states. For mixed RD-cases, as we increase the electric fields, the spin hot spot is shown to cover a wide range of magnetic field. Further, at low magnetic fields or small QD-radii, non-linearity in the phonon mediated spin-relaxation rate for the case of pure Dresselhaus coupling can be found (Figs.\,\ref{fig3GaAs}(a,b), \ref{fig4InAs}(a,b), \ref{fig6-InSb}(a,b) and \ref{fig7}(a,b,c)) due to the interaction of phonons with the heavy-hole bands associated to $|1,2,\pm \rangle$ and $|0,0,\mp \rangle$. Finally, in Figs.~\ref{fig8} and \ref{fig9}, we chose the control parameters (electric fields, magnetic fields, anisotropy and size) of the dots in such a way that heavy-hole  spin-relaxation rate spans within the experimental studies of Ref.~\cite{amasha08}. This may encourage experiments to detect heavy-hole spins in the laboratory in order to test current predictions.

\section{Acknowledgments}\label{acknowledgments}

The simulations were performed at BARTIK High-Performance cluster (National Science Foundation, Grant No. CNS-1624416, USA) in Northwest Missouri State University. S.P. acknowledges Northwest Missouri State University and the Department of Natural Sciences for purchasing COMSOL multiscale multiphysics simulations software package. H.S.C. acknowledges the support of US National Science Foundation Grant No. PHY-2110318. R.M. acknowledges the support of NSERC Discovery and CRC Programs.

\appendix
\section{Spin relaxation for pure Rashba spin-orbit coupling}

\begin{figure}
\includegraphics[width=8.5cm,height=14cm]{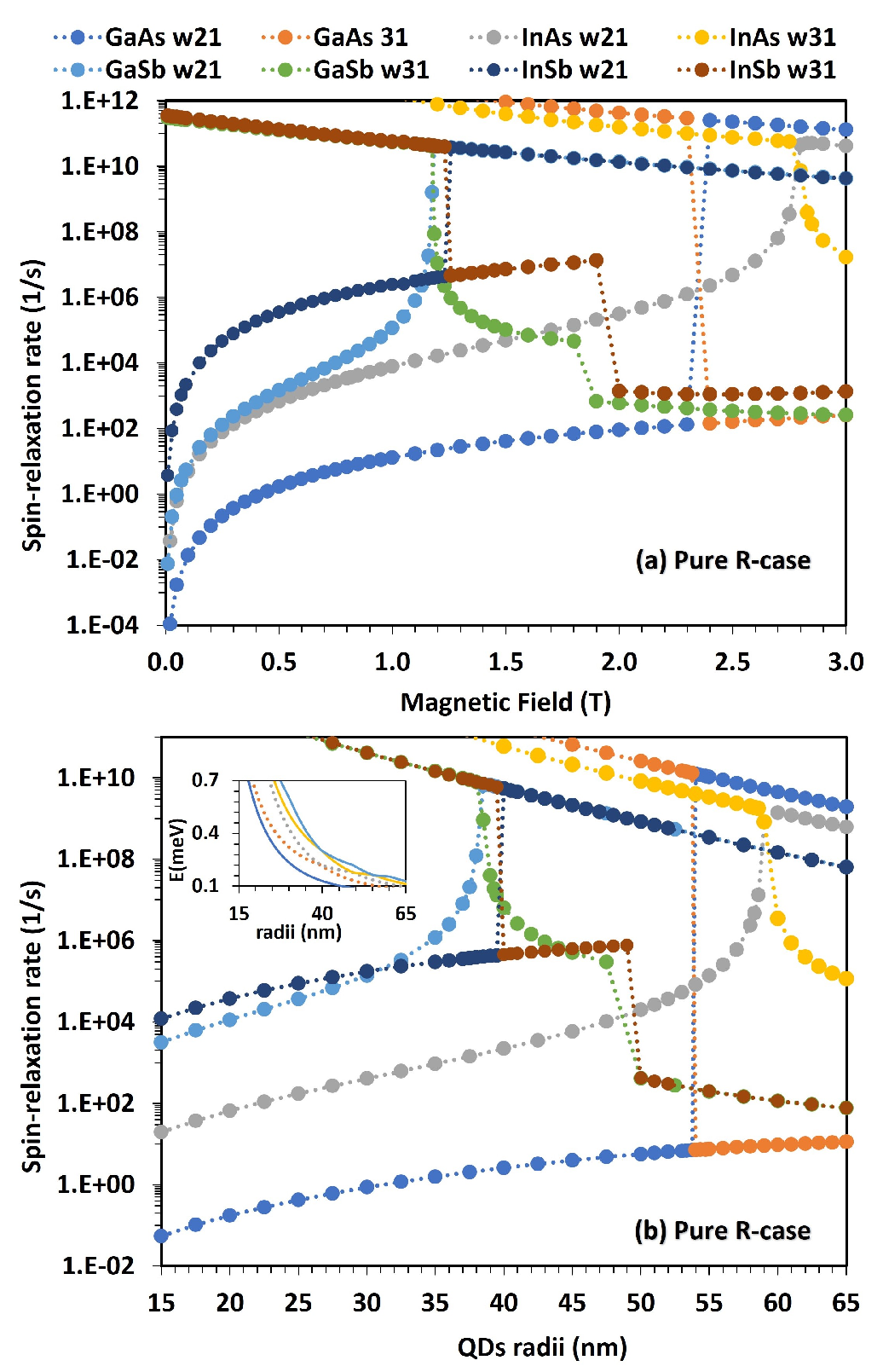}
\caption{\label{figA} Phonon mediated spin-relaxation rate vs the magnetic field (a) and the QDs radius (b) for the pure Rashba spin-orbit coupling case. We chose $\ell_0=25$, $E =10^4$ V/cm (a) and B=0.5T, E =$10^4$ V/cm (b). Notice that the spin hot spot is present for InAs and GaSb dots and it is absent for GaAs and InSb dots.}
\end{figure}

In Figs.~\ref{fig3GaAs}(a) for GaAs, \ref{fig6-InSb}(a) for InSb and \ref{fig7}(a) for GaAs and InSb, we observed that the spin hot spot is present for the pure Dresselhaus spin-orbit coupling and also present for the mixed RD-case. This does not, however, guarantee that the Rashba coupling is able to induce spin-hot spot. For this reason, we have plotted the computed spin-relaxation rate with magnetic fields and dot sizes for the pure Rashba coupling in Fig.~\ref{figA} to confirm that the spin-relaxation rate is a monotonous function of the magnetic fields and dot sizes. In Figs.~\ref{fig4InAs}(a) for InAs, \ref{fig5-GaSb}(a) for GaSb and \ref{fig7}(a) for InAs and GaSb, we observed that spin-hot spot is absent (spin-relaxation rate is a monotonous function of magnetic fields and dots' size) for the pure Dresselhaus coupling case but present for the mixed RD-case. Hence the Rashba spin-orbit coupling must be responsible for inducing the spin hot spot in these dots. To confirm expectation, we have plotted the spin-relaxation rate with magnetic fields and dot sizes in Fig.~\ref{figA} and found that the spin hot spot in the spin-relaxation rate transpires for the pure Rashba spin-orbit coupling. We recall from our main results, for anisotropic QDs, the spin hot spot is found to be present, universally, for both Rashba and Dresselhaus cases.

The inset plot of Fig.~\ref{figA}(b), for the pure Rashba case of InSb, features the band energies as a function of the dot's lateral size. Here, we clearly detect the band crossing (see dotted lines) that, however, may or may not induce spin hot spot depending on the strong or weak admixture of the spin states. The band crossing, as we change the lateral size of the QDs, can also be observed in other materials.

\begin{figure}
\includegraphics[width=8.5cm,height=3.5cm]{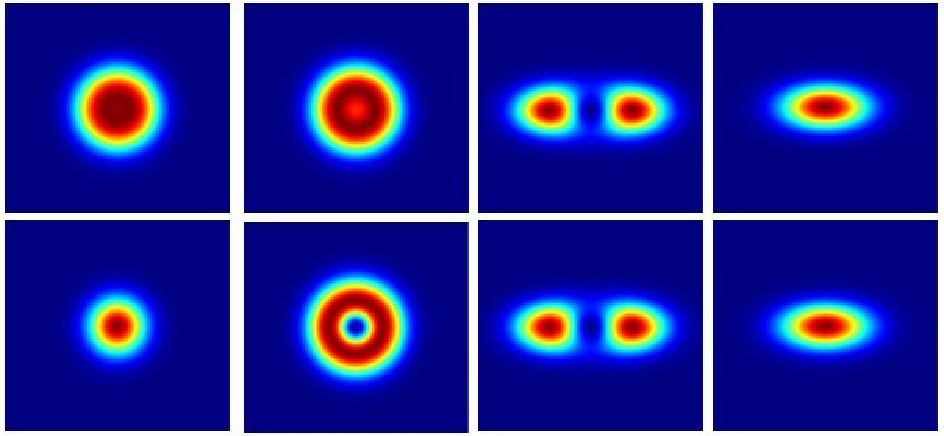}
\caption{\label{figB} The probability density for Pure D-case (first row) and pure R-case (2nd row) for symmetric dots (a=b=1, first and 2nd column) and asymmetric QDs (a=0.25, b=4, 3rd and forth column). Note that the choice of such anisotropy parameters keeps the area of the symmetric and asymmetric dots the same, where spin hot spot can be observed at around the magnetic field of B=2.3T. We chose $\ell_0=25$ and  E =$10^4$ V/cm. }
\end{figure}

We plot the probability density of the spin states $|00-\rangle$ and $|01+\rangle$ for the GaAs QD for the pure Dresselhaus case in Fig.~\ref{figB} (first two figures on first row) and for the pure Rashba case (first two figures on second row), both for an isotropic dot. The identical probability densities for the pure Dresselhaus case confirm that the Dresselhaus spin-orbit coupling is responsible for inducing spin hot spot. In contrast, we find clearly distinct probability densities for the pure Rashba case pointing that the Rashba spin-orbit coupling is not responsible for inducing spin hot spot.

We present, for an anisotropic dot, the probability density of the spin states $|00-\rangle$ and $|01+\rangle$ for the GaAs QD for the pure Dresselhaus case  in Fig.~\ref{figB} (last two figures on first row) and for the pure Rashba case (last two figures on second row). We keep the anisotropic parameters (a=0.25, b=4) in such a way that the area of the dot is identical to the isotropic dot. The clear visual distinctions among the densities confirm that the level crossing of the bands between the states  $|00-\rangle$ and $|01+\rangle$  occurs at smaller values of the magnetic field for anisotropic dots. Similar results are presented for electron QDs in Fig.\,4 of Ref.~\cite{prabhakar11}.

\bibliography{paper23_Bib}

\end{document}